\documentclass[fleqn,12pt,twoside]{article}
\usepackage{beamerarticle}
\usepackage{espcrc1}

\mode<presentation>{\usetheme{Warsaw}}
\usepackage{pgf,pgfarrows,pgfnodes,pgfautomata,pgfheaps,pgfshade}

\usepackage[english]{babel}
\usepackage[latin1]{inputenc}
\usepackage{xmpmulti}

\setbeamercovered{transparent}

\usepackage{graphics}
\usepackage{graphicx}
\usepackage{bm}
\usepackage{amsmath}
\usepackage{amssymb}
\begin{document}

\title{\bf Particle-unstable light nuclei 
with a Sturmian approach that preserves the Pauli principle. }
\author{{L. Canton}$^{a}$, K. Amos$^{b}$, S. Karataglidis$^{c}$,
         G.~Pisent$^{a}$, J.~P.~Svenne$^{d}$, D. van der Knijff$^e$
\\ \ \small
% 
%	 }
%\affiliation{
\\ $^{a}$ Istituto  Nazionale  di  Fisica  Nucleare,
   e  Dipartimento di Fisica  dell'Universit\`a, \\
   via Marzolo 8, Padova I-35131, Italia,\\
   $^{b}$ School of Physics, The University of Melbourne,
   Victoria 3010, Australia\\
   $^{c}$ Department of Physics and Electronics,
   Rhodes University, Grahamstown 6140, South Africa\\
   $^{d}$ Department  of  Physics  and Astronomy,
   University  of Manitoba, and WITP, \\ Winnipeg, MB, Canada R3T 2N2\\
   $^{e}$ Advanced Research Computing, Information Div.,
   The University of Melbourne, Australia\\
   }
                                                                                
\date{\today}

\mode<presentation>{\frame{\titlepage}}                                                                                
\mode<article>{\maketitle}

\mode<presentation>{
\title{On the resonance spectra of particle-unstable ...}                                                                               
\author{{\alert{\bf VARENNA}}\ \ \ \ \ \ \ \ \ \ \ \ \ \ \ \ \ \ \ L. Canton et al.}

\section*{outlines}

\subsection{Part I :Sturmians}

\frame{
\mode<presentation>{  \nameslide{outline}}
  \frametitle{Sturmians}
  \tableofcontents[pausesections,part=1]
}

\subsection{Part II: Coupled-channel Potential and OPP}

\frame{
\mode<presentation>{  \nameslide{outline}}
  \frametitle{Model Coupled-Channel Potential and OPP}
  \tableofcontents[pausesections,part=2]
}

\subsection{Part III: Applications}

\frame{
\mode<presentation>{  \nameslide{outline}}
  \frametitle{Applications}
  \tableofcontents[pausesections,part=3]
}

}
%%%%%%%%%%%%%%%%%%%%%%%%%%%%%%%%%%%%%%%%%%%%%%%%%%%%%%%%%%%%                                                                                
\mode<article>{

\begin{abstract}
Sturmian theory for nucleon-nucleus scattering is discussed in 
the presence of all the phenomenological
ingredients necessary for the description of 
weakly-bound (or particle-unstable) light nuclear systems. 
Currently, we use a macroscopic potential model of collective nature. 
The analysis shows that the couplings
to low-energy collective-core excitations are fundamental but they 
are physically meaningful only if the constraints introduced by the 
Pauli principle are taken into account. The formalism 
leads one to discuss a new concept, Pauli hindrance, which appears 
to be important to understand the structure of weakly-bound 
and unbound systems.
\end{abstract}
%{\LARGE Never underestimate a nuclear model \\
%that reproduces \alert{bound and scattering} \\
%spectra with the same Hamiltonian ! }
%\end{abstract}
%\pacs{24.10-i;25.40.Dn;25.40.Ny;28.20.Cz}
%}
}
%%%%%%%%%%%%%%%%%%%%%%%%%%%%%%%%%%%%%%%%%%%%%%%%%%%%%%%%%%%%%%%%%%%%%%%%%

\mode<presentation>{\part{Sturmians}

\frame{\partpage}
}

\section{Sturmians}
Sturmians provide the solution of the 
scattering problems by matrix manipulation and represent
an efficient formalism for determination of S-matrix, and scattering 
wave functions, bound states and resonances~\cite{WWS}.
They work well with non-local interactions (such as those non localities
arising from the effects due to Pauli exchanges).
They allow a consistent treatment of Coulomb plus nuclear interactions,
as well as the inclusion in the scattering process of coupled-channel (CC) 
dynamics. This occurs, for instance, when strong coupling to 
low-lying excitations of the target nucleus have to be taken into account. 
%\item Algebraic derivation of the Optical potential (DPP)

\frame{
%Sturmians ({\em also known as} Weinberg states) represent an {\it alternative} 
%way to formulate the Quantum Mechanical problem.

Consider the Schr{\"o}dinger equation
$(\alert{E}-H_o)\Psi_{\alert{E}} = V \Psi_{\alert{E}}$ in the standard 
(time-independent)  way
%\begin{equation}
%$(\alert{E}-H_o)\Psi_{\alert{E}} = V \Psi_{\alert{E}}$ ,
%\end{equation}
where \alert{E} is the spectral variable, and $\Psi_{\alert{E}}$ 
is the eigenstate.
Sturmians, instead, are the eigensolutions of:
\begin{equation}
(E-H_o)\Phi_{\alert{i}}(E) = \frac{V}{\alert{\eta_i}(E)} \Phi_{\alert{i}}(E)\, ,
\end{equation}
where E is a parameter. The eigenvalue \alert{$\eta_i$} is the potential scale.
Thus the spectrum consists of all the potential rescalings that give 
solution to that equation, for given energy E, 
and with well-defined boundary conditions.

%Standard boundary conditions of $\Phi_i(E)$:
%\begin{itemize}
%\item[{$E < 0$}] Bound-state like; normalizable.
%\item[{$E > 0$}] Purely outgoing/radiating waves; non-normalizable.
%\end{itemize}
}

\frame{
%For any energy $E$, Sturmians are \alert{$V$}-complete and 
%\alert{$V$}-normalizable:
%\begin{equation}
%V =\sum_i V|\Phi_i\rangle %{1\over\eta_i} 
%\langle \Phi_i|V
%\end{equation}
%\begin{equation}
%%\eta_i
%\delta_{ij}=\langle \Phi_i|V|\Phi_i\rangle
%\end{equation} 
%(\alert{Careful! the normalization is non trivial})

The standard boundary conditions for Sturmians $\Phi_i(E)$ are
bound-state like normalizable for $E<0$, while for {$E > 0$} 
these states are purely outgoing/radiating waves, and non-normalizable.
Their spectrum is purely discrete and bounded absolutely. 
For short-range (nuclear-type) potentials, the eigenvalues can accumulate
around 0 only.
}

\frame{
%Then, the single-channel $S$-matrix can be written as
%\begin{equation}
%S(E)=\frac{\Pi_i (1-\eta_i(E^{(-)}))}{ \Pi_i (1-\eta_i(E^{(+)}))}
%\end{equation}
Introducing the factor $\hat\chi_{ci}(E,k) = <k,c| V | \Phi_i(E)>$ in momentum
space, 
the CC $S$-matrix can be written as
\begin{equation}
S_{cc'}(E)= \delta_{cc'} - i \pi \sqrt{k_c k_c'} 
\sum_i \hat\chi_{ci}(E^{(+)};k_c)\frac{1}{ 1-\eta_i(E^{(+)})} 
\hat\chi_{c'i}(E^{(+)};k_{c'})
\end{equation}

%We interpret le construction as follows: 
%the scattering process initiated in the asymptotic channel 
%$c$  is ``captured'' into Sturmians. Subsequently the Sturmian propagates 
%freely in the interaction region, and finally decays into the outgoing 
%hannels. 
This structure naturally reflects the description of 
the scattering process in terms of compound nucleus formation,
and leads to an expression which is rather similar to that obtained
in $R$-matrix formalism. 

%\subsection{Resonances and bound states in terms of Sturmians}

Resonant structures as well as bound states can be obtained 
in terms of the properties of Sturmian eigenvalues.
A bound state occurs when one of the eigenvalues moves toward
the right on the real axis, and crosses the value 1 at some negative 
energy. That particular energy value corresponds to the 
bound-state energy. On the other hand, a resonance is found if any eigenvalue,
initially varying along the real axis, becomes complex before reaching
the value 1. The resonance energy then is that value (positive since
the scattering threshold has been reached) for which the real part
of the complex eigenvalue ($\eta_i(E)$) equals 1.
The width of the resonance can also be determined 
geometrically by the patterns 
of the sturmian trajectories, and relates to the imaginary part
of the sturmian eigenvalue at the resonant energy. 
%Such patterns can be seen in Fig.~\ref{fig1}. 
%In Fig.~\ref{fig2} one observes the situation in a
%realistic case, namely $n$-$^{12}C$ elastic scattering. The eigenvalue
%trajectories produce two $3/2^{+}$ resonances that can be clearly seen
%in the experimental data as well as in the theoretical calculation.

%\frame{
%\begin{figure}[ht]

%\scalebox{0.35}{\includegraphics{1.5plus.eps}} 
%\scalebox{0.58}{\includegraphics{n-C12-sig-lin-1.5plus.eps}}
%\caption{\label{fig2}
%Neutron-$^{12}$C in the 3/2$^+$ channel: a realistic 
%case. Low-energy resonances in 3/2$^+$ $n$-$^{12}$C 
%system. Sturmian patterns (left) and 3/2$^+$ resonant cross-section (right).
%The dashed line in the left panel denotes the unit circle.}
%\end{figure}
%}
%\caption{\label{Fig2}
%}
%\end{figure}

\mode<presentation>{\part{Collective Coupled-Channel potential and OPP}
\frame{\partpage}
}

\section{Model Coupled-channel potential and OPP}

\frame{
In the current description, we adopt the macroscopic potential approach:
a nucleon is scattered by a nucleus (light nuclei with 0$^+$ g.s. are 
considered) and we include couplings to first core excitations 
of collective nature (quadrupole, octupole, etc.) since these couplings 
play an important role in the dynamics. 
%Detailed expressions
%of the various terms that have been included in the potential
%have been published in Ref.~\cite{Amos} and subsequent publications.

\subsection{First application: n-$^{12}$C}
\frame{
As a first application, we considered the scattering of
neutrons off $^{12}$C, coupling the ground state of the target
to  the first two low-lying 
excitations $2_1^+$ (4.43 MeV) $0_2^+$ 
(7.63 MeV), and searched parameters to obtain a description
of the resonant spectra and scattering cross 
sections~\cite{Amos,Canton,Pisent}.
However, many deeply-bound spurious states occurred
in the bound spectrum, and it was not possible to obtain a consistent
description of both bound structure and scattering data with 
the same CC Hamiltonian. These spurious
deeply-bound states originate from violation of 
the Pauli principle. 
%The phase-space corresponding to 
%target nucleons in fully occupied shells has to be inhibited 
%to the incoming nucleon, but in the original CC model such condition 
%was missing. 

Various methods have been suggested to remove the deeply-bound
forbidden states.
% A recent article~\cite{Brink} on the subject
%contains an historical review on the various approaches and their
%connections.
%In phenomenological macroscopic-type calculations,
%such Pauli condition has been implemented first by 
%introducing the Orthogonality
%Condition Model~\cite{Saito}. Alternatively, the orthogonality condition
%can be introduced directly in the Hamiltonian by adding a new term
%in the potential, the highly non-local Orthogonalizing 
%Pseudo-Potential~\cite{Kukulin}. 
%With the advent of super-symmetric
%quantum mechanics\cite{Witten}, it is now possible to define super-symmetric
%transformations\cite{Baye} that produce new local (and highly singular) 
%potentials which also generate spectra free of spurious states. 
In our approach, we use the technique of Orthogonalizing 
Pseudo Potentials~\cite{Kukulin}, which eliminates the spurious states
by the addition of a new term in the nuclear potential~\cite{Canton}.
}

\frame{
This new term has the form 
(in partial-wave decomposition)
\begin{eqnarray*}
{\cal V}^{OPP}_{cc'}(r,r')=
 \delta_{cc'} 
\lambda_{c} 
A_{c}(r)
A_{c}(r') (\delta_{c=s\frac{1}{2}})
+ \delta_{cc'}
\lambda_{c} 
A_{c}(r)
A_{c}(r')  (\delta_{c=p\frac{3}{2}}) \, ,
\end{eqnarray*}
where $A_c(r)$ are the 
\alert{Pauli-forbidden} deep (CC-uncoupled) bound states.
The use of large positive values for the $\lambda$ parameter 
eliminates the deeply-bound spurious states.
%, , since  
%{a state in the OPP approach is }
%{\alert{\em forbidden} in the  limit $\lambda\rightarrow +\infty$},
%while it is
%{{\em allowed}  when $\lambda\rightarrow 0$}
}

\frame{
%In $^{13}C$, two shells  are forbidden to the spare nucleon:\\  
%$s\frac{1}{2}$ and $p\frac{3}{2}$.

%\begin{figure}[b]

%\centerline{Bound-state levels and total cross-sect}
%\centerline{\scalebox{0.7}{\includegraphics{C13-levels-OPP-fig.eps}}
%\scalebox{0.3}{\includegraphics{n-c12-sect.eps}}}
%\caption{\label{Fig5}

%\end{figure}

%\centerline{Total Cross-section at low energy}
}

%\subsection{The {${\lambda}$}-dependence}

\setbeamercovered{covered}
\begin{frame}
%\centerline{The $\lambda$ dependence 
% \alert{\only<1>{100}
%\only<2>{200}\only<3>{300}\only<4>{500}\only<5>{1000}\only<6>{5000}}}
%\animate<2-4>
%\scalebox{0.6}{\multiinclude[format=eps]{n-C12-sig-log-lambda}}

\begin{figure}

\scalebox{0.5}{
\includegraphics{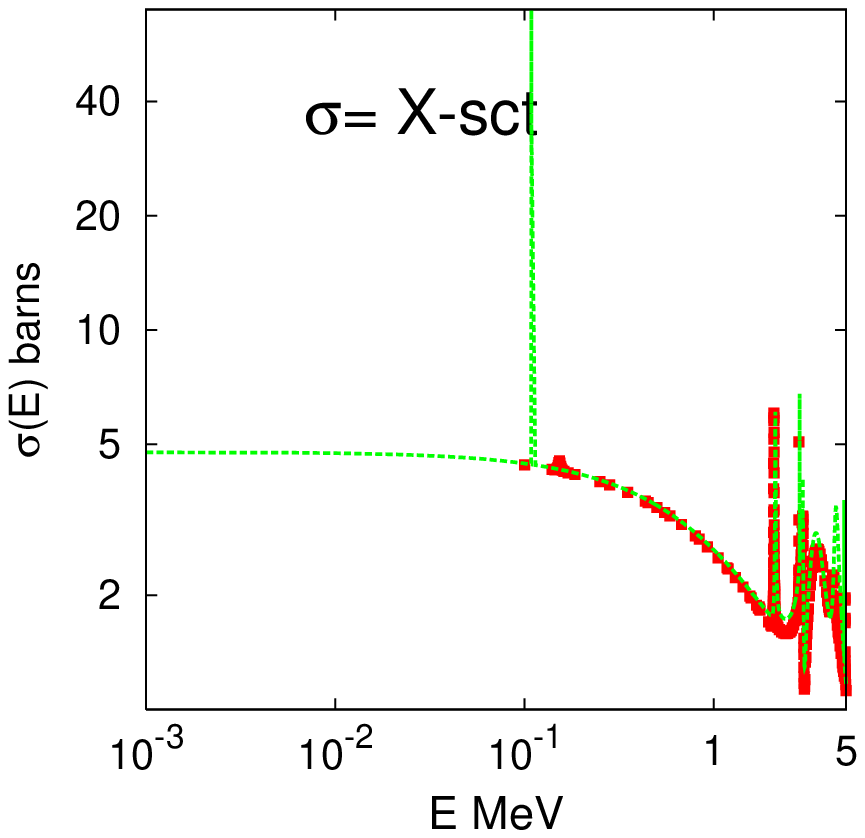}\hskip -2.0truecm
%}
%
%\scalebox{0.5}{
\includegraphics{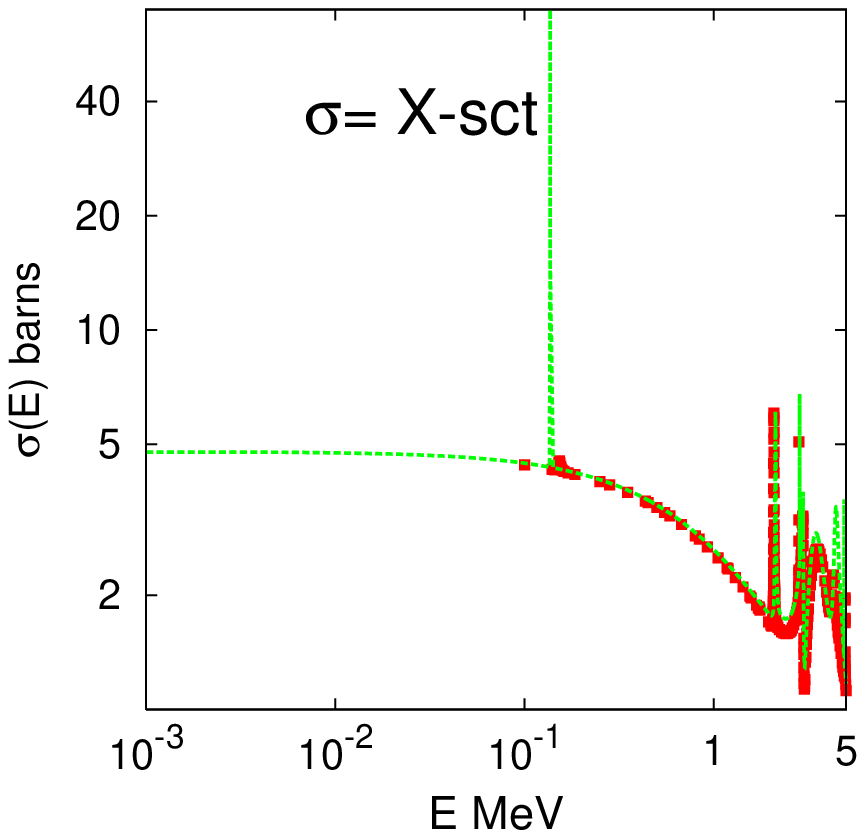} \hskip -2.0 truecm,
\includegraphics{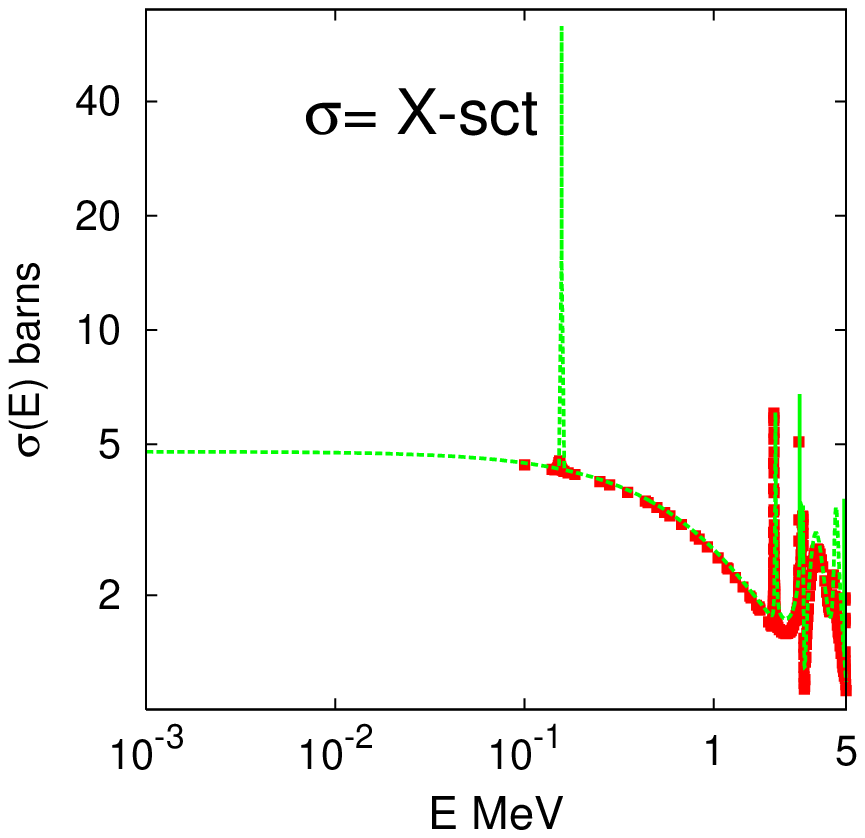}}
\caption{\label{Fig3}
The $\lambda$ dependence (of the elastic cross section
for n-$^{12}$C scattering) with values 
300, 500, and 1000 MeV (left, center, right, respectively).
Observe the stability of the 5/2$^-$ resonance at 0.1 MeV.}
\end{figure}
%\multiinclude{n-C12-sig-log-lambda}

A key feature of the OPP method is the effects on results 
with variation in $\lambda$. Three cases are shown in
Fig.~\ref{Fig3} and stability is reached using $\lambda= 1000$ 
MeV.

\end{frame}

%\section{Applications of MCAS theory}

%\subsection{Analyzing powers of nucleons off $^{12}C$}

\frame{

%\centerline{Analyzing powers of nucleons off $^{12}C$}

%Spectral analyses of n-$^{12}$C system were published in
%Refs.~\cite{Amos,Canton,Pisent}.
%Note that all the parameters, including the spin-orbit term, have 
%been fixed on the known spectrum of $^{13}C$. 
Shortly after our analyses were completed, 
new spin-polarization data measured at TUNL were published~\cite{TUNL}.
Without any parameter adjustments or tuning, we could reproduce 
those low-energy data~\cite{Svenne}. That tested  the
spin-orbit parametrization and validated its spin-structure.

\subsection{Low-lying unbound states of $^{15}$C and $^{15}$F}

A second analysis concerned an unbound nucleus, $^{15}$F, whose
properties were studied in connection with its weakly-bound partner $^{15}$C. 
Our analysis was triggered by recent low-energy 
experimental data on the $^{14}$O-proton system, in inverse kinematics.
We included in the model low-lying excitations of $^{14}$O/$^{14}$C, 
in a macroscopic CC model with parameters given in 
Ref.~\cite{Canton06}.

{\begin{figure}

\centerline{\includegraphics[height=8.cm]{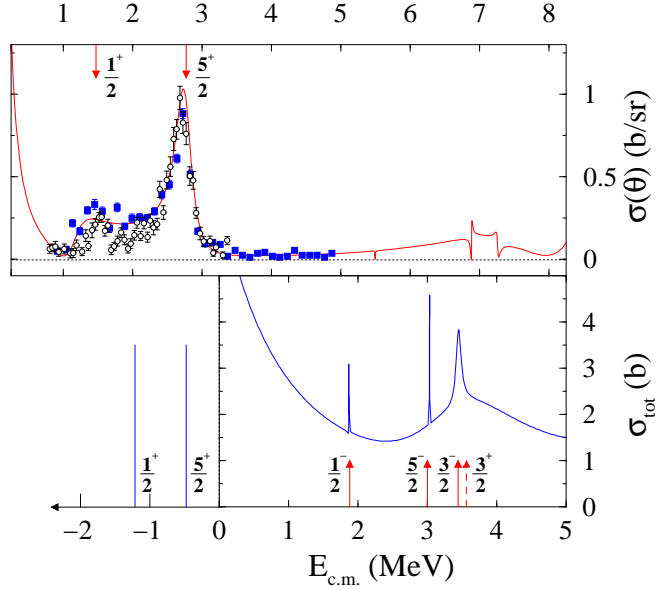}} 
%{\includegraphics[height=8.5cm]{Fig1-fin.eps}}
\caption{\label{Fig5}
$^{15}$F  (top, $p-^{14}$O differential cross-section\@ $\theta_{cm}=180^o$) 
\& $^{15}$C (bottom, the bound spectrum and the elastic 
scattering cross-section).}
\end{figure}
%\end{beamercolorbox}
}

We used known properties of $^{15}$C to predict (see Fig.~\ref{Fig5}) 
new states in 
$^{15}$F, in particular three narrow resonances of negative parity, 
1/2$^-$, 5/2$^-$ and 3/2$^-$, in the range 5-8 MeV. 
To obtain those results, a new concept had to be introduced~\cite{Canton06}:
it is the concept of Pauli-hindered states. Up to now
we have discussed states that are Pauli allowed ($\lambda\simeq 0$)
or Pauli forbidden ($\lambda\ge 1$ GeV). Now we introduce
a state that sits in between: a state that is neither prohibited 
nor allowed but simply suppressed, or {\em hindered}, by the Pauli principle.
We expect that this situation can apply in weakly-bound/unbound 
light systems, where the formation of shells may become critical.
These hindered states can be conveniently described in the OPP
scheme with values of $\lambda$ of a few MeV.
The need to consider intermediate situations between 
Pauli blocking and Pauli allowance has been registered before in the
literature, mostly in connection with RGM approaches~\cite{Langanke}.

\subsection{P-shells in mass=7 nuclei}

We consider the spectral structure the 
mass=7 isobars: $^7$He, $^7$Li, $^7$Be, and $^7$B. We describe 
these systems in terms of nucleon-nucleus interaction, explicitly
including the low-lying core excitations of the mass-6 sub-system.
Thus we determine the bound and resonant
(above nucleon emission) spectra starting from a nucleon-$^6$He collective 
CC interaction model coupling the $g.s.$ of $^6$He
to the first and second $2^+$ excitations (for $^7$He, $^7$Li).
Similarly, for $^7$Be, and $^7$B, we use $^6$Be as the core system.
This description of the mass-7 systems is in many ways alternative
to those based on cluster models ($^7$Li as a $^3H$+$\alpha$ dicluster, 
etc.) or to those based on more microscopic models, either no-core 
shell model calculations or Green Function Monte Carlo 
calculation. A comparison amongst different descriptions (dicluster model,
microscopic shell model, and collective-coupling model)
for mass-7 nuclei is given in Ref.~\cite{Canton06a}. 
\begin{table}

\begin{columns}
\begin{column}[T]{5.0cm}
%\begin{center}
\scalebox{0.6}{\begin{tabular}{c|cc|cc}
  $J^\pi$  & \multicolumn{2}{c}{${}^7$Li} & \multicolumn{2}{c}{${}^7$Be}\\ 
 \hline                                                  
 & Exp. & Theory & Exp. & Theory \\
  \hline
$\frac{3}{2}^-$   & $-$9.975         & $-$9.975      &    $-$10.676          & $-$11.046       \\
$\frac{1}{2}^-$   & $-$9.497         & $-$9.497      &    $-$10.246          & $-$10.680       \\
$\frac{7}{2}^-$   & $-$5.323 [0.069] & $-$5.323      &    $-$6.106 [0.175]   & $-$6.409        \\
$\frac{5}{2}^-$   & $-$3.371 [0.918] & $-$3.371      &    $-$3.946 [1.2]     & $-$4.497        \\
$\frac{5}{2}^-$   & $-$2.251 [0.08]  & $-$0.321      &    $-$3.466 [0.4]     & $-$1.597        \\
$\frac{3}{2}^-$   & $-$1.225 [4.712] & $-$2.244      &          $--$         &    $--$         \\
$\frac{1}{2}^-$   & $-$0.885 [2.752] & $-$0.885      &                       & $-$2.116        \\
$\frac{7}{2}^-$   & $-$0.405 [0.437] & $-$0.405      &    $-$1.406 [?]       & $-$1.704        \\
$\frac{3}{2}^-$   &      $--$        &    $--$       &    $-$0.776 [1.8]     & $-$3.346        \\
$\frac{3}{2}^-$   &    1.265 (0.26)  & 0.704 (0.056) &     0.334 (0.32)       & $-$0.539       \\
$\frac{1}{2}^-$   &                  & 1.796 (1.57) &                     &  0.727 (0.699)     \\
$\frac{3}{2}^-$   &  3.7 (0.8) ?$^a$ & 2.981 (0.99)&                 & 1.995 (0.231)   \\
$\frac{5}{2}^-$   &  4.7 (0.7) ?$^a$ & 3.046 (0.75)&                  & 2.009 (0.203)   \\
$\frac{5}{2}^-$   &                  & 5.964 (0.23)  &                       & 4.904 (0.150)   \\
$\frac{7}{2}^-$   &                  & 6.76  (2.24)  &      6.5 (6.5) ?$^a$  & 5.78  (1.65)   \\
\hline
\end{tabular}}				                          
% \scalebox{0.4}{$^a$For these states spin and parity are unknown}
 %\begin{center}					                          
 %\end{center}					                          
% \scalebox{0.5}{$^b$Spin-parity of this state has been assigned as $\frac{1}{2}^-$}
%\end{center}					                          
%\begin{center}
\end{column}
\begin{column}[T]{5.0cm}
\scalebox{0.6}{\begin{tabular}{c|cc|cc}
$J^\pi$  & \multicolumn{2}{c}{${}^7$He} & \multicolumn{2}{c}{${}^7$B} \\
\hline                                                  
 & Exp. & Theory & Exp. & Theory\\
\hline
$\frac{3}{2}^-$   & 0.445 (0.15)     & 0.43 (0.1)  &    2.21 (1.4)    & 2.10 (0.19)       \\
$\frac{7}{2}^-$   &  $--$            & 1.70 (0.03) &                & 3.01 (0.11)     \\
$\frac{1}{2}^-$   &  1.0 (0.75) ?$^a$ & 2.79 (4.1)  &             & 5.40 (7.2)      \\
$\frac{5}{2}^-$   & 3.35 (1.99)     & 3.55 (0.2)  &                & 5.35 (0.34)      \\
$\frac{3}{2}^-$   & 6.24 (4.0) ?$^a$  & 6.24 (1.9)  &                     &                  \\
\hline
\end{tabular}}				                          
%\scalebox{0.5}{$^a$ Observed very recently and interpreted as a $\frac{1}{2}^-$ state.} 
%\scalebox{0.5}{$^b$ Spin-parity of this state is unknown.}
\end{column}
\end{columns}
\caption{\label{table-mass-7} \small 
Results for $^7$Li and $^7$Be states (left table), and for $^7$He and $^7$B 
(right table). With widths in brackets, energies are in MeV and relate to 
scattering thresholds for  n-${}^6$He or n-${}^6$Be. 
For states labeled by ``?$^a$'' spin-parity attributes are
unknown or uncertain.}
\end{table}

%%%%%%%%%%%%%%%%%%%%%%%%%%%%%%%%%%%%%%%%%%%%%%%%%%%%%%%%%%%%%%%%%%%%%%%%%%%%%%%%%%%%
 In Table~\ref{table-mass-7} we compare results with experiments~\cite{Canton06a}.
 A single CC potential model has been used for all four nuclides.
 The results for $^7$Li/$^7$Be (and for $^7$He/$^7$B) differ solely
 by the effect of the central Coulomb field. Instead, if we compare
 results for the pairs  $^7$Li/$^7$He and  $^7$Be/$^7$B, they differ
 solely for the different action of the OPP term.
 For the two mass-7 bound systems, Pauli blocking is assumed in
 the $0s_{1/2}$ shells and all the remaining shells are 
 considered allowed, while for the two unbound systems 
 a more complex OPP scenario is considered: 
 the $0s_{1/2}$ shells are blocked, the $p$ shells are hindered,
 and only the higher shells are completely allowed. This hindrance
 of the $p$ shells could reflect the anomalous interaction
 of the neutron with $^6$He, which is a neutron halo. A similar situation
 could occur in the mirror case of $^7$B, with the interaction of a proton 
 with a $^6$Be-type core. In our calculation, we have made the hypothesis
 of a Pauli hindrance in the p-shells defined as follows:
 $(0p_{3/2}[0^+_1;2^+_1;2^+_2])$, 
 $(0p_{1/2}[0^+_1,2^+_1;2^+_2])$. The extended nature 
 of the even-even mass-6 subsystems, 
 either weakly-bound ($^6$He) or unbound ($^6$Be), is approximately 
 reflected also in the geometric size of the potential parameters 
 with rather extended radius, diffuseness, and quadrupole deformation.
 
 \section{Conclusions}
 A Sturmian-based approach has been applied
 to CC problems at low energy, using phenomenological
 potentials with macroscopic, collective-type couplings. 
 But the method is sufficiently flexible that it could be applied also to
 microscopically generated potentials.
 The approach has been applied to stable nuclei, as well as to weakly-bound and to unbound 
 (with respect to the nucleon emission threshold)
 light nuclei. In the few cases considered, interesting results 
 have been obtained, sometimes with rather good
 reproduction of bound spectra and scattering observables, and often with 
 predictions that could stimulate new experiments. In this approach,
 the highly nonlocal OPP term is crucial in order to include macroscopically
 the effects of Pauli exchanges. Finally, for weakly-bound or unbound light 
 systems, the concept of Pauli hindrance is suggested.
 It implies that the nuclei have partially forbidden $p$-shells,
 namely ``proto''-shells where the exclusion principle
 neither forbids nor allows occupancy. Access to the relevant phase-space
 is, to some extent, simply suppressed.

%\begin{frame}
%\centerline{States for $^7$He and $^7$Li with the same nuclear potential}
%\centerline{Only the OPP term changes: 
%\only<1>{\alert{for $^7$Li}}\only<2>{\alert{for $^7$He}}}
%{\bf Pauli Forbidden $\lambda\simeq 1 GeV$}
%\begin{eqnarray*}
%\alert{0s_{1/2} + 0^+_1} & \alert{0s_{1/2} + 2^+_1} & \alert{0s_{1/2} +
%2^+_2}  \\
%\end{eqnarray*}
%\only<1>{\bf Pauli Allowed $\lambda\simeq 0 MeV$}
%\only<2>{\bf Pauli Hindered $\lambda\simeq 1\div 10 MeV$}
%\only<1>{\begin{eqnarray*}
%{0p_{3/2} + 0^+_1} & {0p_{3/2} + 2^+_1} & {1p_{3/2} +2^+_2} \\
%{0p_{1/2} + 0^+_1} & {0p_{1/2} + 2^+_1} & {0p_{1/2} +2^+_2}
%\end{eqnarray*}
%$\ $\\
%$\ $\\
%$\ $}
%\only<2>{\begin{eqnarray*}
%\alert{0p_{3/2} + 0^+_1} &\alert{0p_{3/2} + 2^+_1}&\alert{1p_{3/2} +2^+_2} \\
%\alert{0p_{1/2} + 0^+_1} &\alert{0p_{1/2} + 2^+_1}&\alert{0p_{1/2} + 2^+_2}\\
%\end{eqnarray*}
%$\lambda(0p_{3/2}[0^+_1;2^+_1;2^+_2])=17.6 MeV$, \\
%$\lambda(0p_{1/2}[0^+_1])=36.0 MeV$, \\
%$\lambda(0p_{1/2}[2^+_1;2^+_2])=5.6 MeV$
%}

%\end{frame}

\end{document}